# Bayesian hierarchical reconstruction of protein profiles including a digestion model

**Pierre GRANGEAT[1], Pascal SZACHERSKI[1,2], Laurent GERFAULT[1], Jean-François GIOVANNELLI[2]**

[1] CEA, LETI, MINATEC Campus, DTBS, 17 rue des Martyrs, F-38054 Grenoble cedex 9, France.
[2] Université de Bordeaux 1 – CNRS - IPB, IMS, 351 Cours de la Libération, F-33405, Talence cedex, France.

**Introduction** :
Mass spectrometry approaches are very attractive to detect protein panels in a sensitive and high speed way. MS can be coupled to many proteomic separation techniques. However, controlling technological variability on these analytical chains is a critical point. Adequate information processing is mandatory for data analysis to take into account the complexity of the analysed mixture, to improve the measurement reliability and to make the technology user friendly. Therefore we develop a hierarchical parametric probabilistic model of the LC-MS analytical chain including the technological variability. We introduce a Bayesian reconstruction methodology to recover the protein biomarkers content in a robust way. We will focus on the digestion step since it brings a major contribution to technological variability.

**Method** :
In this communication, we introduce a hierarchical model of the LC-MS analytical chain. Such a chain is a cascade of molecular events depicted by a graph structure, each node being associated to a molecular state such as protein, peptide and ion and each branch to a molecular processing such as digestion, ionisation and LC-MS separation. This molecular graph defines a hierarchical mixture model. We extend the Bayesian statistical framework we have introduced previously [1] to this hierarchical description. As an example, we will consider the digestion step. We describe the digestion process on a pair of peptides within the targeted protein as a Bernoulli random process associated with a cleavage probability controlled by the digestion kinetic law.

**Preliminary data :**
In order to produce simulated data, we consider the experimental framework described in [2] applied to human NSE combining immunoenrichment, trypsin digestion and LC-MS analysis on an Orbitrap. PSAQ quantification is performed spiking an isotopically-labeled version of NSE [3]. We define 5 amounts of unlabeled NSE from 0 to 400 ng/ml within 1 ml serum samples. We analyze 2 peptides assuming artificially that there exists a miscleavage delivering a molecule combining both peptides.
Technological variability is introduced by considering variation of 10% on LC retention time and 30% on MS ionisation gain. Variability on digestion is induced in particular by temperature and pH [4]. We model the digestion coefficient linking protein and peptides content as a constant yield weighted by the Bernoulli parameter p for the cleaved peptides and parameter 1-p for the miscleaved molecule. We make this parameter varying between 0.4 and 1, but we assume its value to be known.

Two data processing strategies are studied. Firstly, the digestion variability is ignored and the protein content is estimated from the 2 cleaved peptides. Secondly, the miscleaved molecule is introduced and the digestion variability is monitored. Parameters are estimated within the Bayesian hierarchical reconstruction. 321 spectrograms are simulated in order to compare the performances of the two strategies. Coefficient of variation of estimated concentrations is computed for first (CV1) and second strategy (CV2).

The ratio CV2/CV1 is respectively 0.8, 0.7, 0.7, 0.2, 0.05 for concentrations of 0, 50, 100, 200 and 400 ng/ml, demonstrating a significant improvement for the largest concentrations.

**Novel aspect** :
Bayesian hierarchical inversion, technological variability, digestion model, robust protein quantification.